# Flowing Bottomhole Pressure Calculation for a Pumped Well under Multiphase Flow.

*Authors: Bikbulatov S., Khasanov M., Zagurenko A.*

## Summary

The ability to monitore bottomhole flowing pressure in pumping oil wells provides important information regarding both reservoir and artificial lift performance. Converting surface pressure measurements to bottomhole is currently accomplished by locating the fluid level in the annulus using a sonic device and then applying a correlation to estimate the density of the gas-cut liquid column above the perforations.

This work proposes a flowing bottomhole pressure calculation procedure from fluid level measurements. The model is developed from experimental work and from theoretical arguments.

The calculation procedure developed allows to calculate BHP without shutting the well, which is common for fluid level casinghead pressure measurements. Also this method allows to take into account real geometry of the well.

The comparison of the calculated and measured pump intake pressure shows good accuracy of a technique. It allows to draw a conclusion for an opportunity to use this method in practice.

## Introduction

*"Is the well producing all the fluid that it is capable of producing without problems?"*
*A. L. Podio*

The producing-rate efficiency of a well can be determined with the curve of inflow performance relationship[1], which requires knowledge of the BHP.

The ability to monitor bottomhole pressures provides many advantages for reservoir management. Pumping wells completed without a packer provide a special opportunity for this low cost and reliable bottomhole pressure surveillance. For these pumping wells, wellhead pressure data is converted to bottomhole pressure by use of flow models and an acoustic (sonic) device to locate the gas-liquid interface. Fig. 1 shows the schematic of this type of pumping oil well. The well is completed in a conventional fashion, without a packer. The pump can be a sucker rod pump, PC pump or ESP. The produced fluids are pumped from the well through the tubing string, while produced and solution gas travels up the tubing/casing annulus and is produced as casinghead gas at the surface. Acoustic devises are used to determine the depth to the gas-liquid interface. Once it has been located, bottomhole pressure is estimated through use of flow models to calculate the pressure drop through the gas phase above the interface and oil-water-gas mixture that exists below the interface.

From the knowledge of the lengths of gas and liquid columns, BHP can be estimated by adding the pressures exerted by these columns to the casinghead pressure. Although simple in concept, this indirect calculation presents potential problems in two areas: resolution of the acoustic device measuring the gas/liquid interface and estimation of the gas-entrained-liquid-column density.

Significant progress has been made in the acoustic device's ability to monitor the movement of the liquid column as function of time. Estimation of change in liquid-column density, however, still fraught with uncertainties.

The fluid distribution in the annulus is a function of the producing conditions of the particular well. The situation is generally found in the field: the liquid level is above the formation and casinghead gas is produced. This condition results in a gaseous annular liquid column. At stabilized producing conditions, the oil in the casing annulus becomes saturated with the gas that is continuously flowing to the surface.

The flow of gas through a static liquid column creates a special type of multiphase flow, termed Zero Net Liquid Flow (ZNLF) (fig.1). In this case, liquid is present in the wellbore but does not flow out the tubing/casing annulus with the gas phase. The gas phase simply passes through the column of liquid. A number of methods have been developed to predict the liquid holdup of this gas-cut liquid column. The well-known Gilbert[2] chart correlates a Liquid Correction

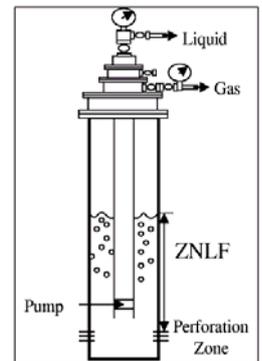

Fig.1. Schematic of Pumping Oil Well

Factor (*LCF*). The *LCF* corrects the liquid density to account for the effect of the gas phase. In modern terms, the *LCF* is related to liquid holdup, $H_L$, by:

$$(LCF)\rho_L = \rho_m = \rho_L H_L + \rho_G(1 - H_L) \quad (1)$$

In 1977, Godbey & Dimon[3] presented a correlation of $v_{sg}$ with the gas void faction, $1 - H_L$. The most widely a use method was proposed by Podio et al.[4] in 1980. Hasan et al.[5] presented a model for prediction of the *LCF* which allows a variety of fluid properties to be considered. In 1994, Kabir & Hasan[6] presented a comprehensive review of the current methods and discussed the accuracy of the various available methods. Research in two areas - mass transfer in gas/liquid systems and two-phase flow - has produced a wealth of information for predicting gas void fractions in stagnant liquid columns. None of these correlations, however, account for the effect of casing and tubing diameters.

The purpose of this project is to explore the relevant literature and develop a model for flowing bottomhole pressure calculation from fluid level measurements using a practical range of flow conditions encountered in a pumping-well annulus.

# 1 Flow model

Various flow models are used for casing and tubing/casing annulus description because of different geometries. Works of Hasan and Kabir[20,21,22] was taken as a basis for this flow models.

Hasan and Kabir also developed a mechanistic model to predict pressure gradients in wellbores. To model flow-pattern transitions, Hasan and Kabir adapted an approach very similar to that of Taitel et al.[23] Hasan and Kabir identified the same four flow patterns: bubble flow, slug flow, chum flow, and annular flow.

## 1.1 Flow model for casing.

**Bubble/Slug Transition.** Transition from bubble flow (the condition of small bubbles dispersed throughout the flow cross section) to slug flow (when the bubble becomes large enough to fill the entire cross section) requires agglomeration or coalescence. Bubbles, other than very small ones, generally follow a zigzag path when rising through a liquid. This results in collisions among bubbles, with the consequent bubble agglomeration and formation of larger bubbles, which

$$v_{Sg} = \frac{\sin\theta}{4-C_o}(C_o v_{SL} + v_s) \quad (2)$$

$C_0$ is the flow coefficient given by Eq. 3

$$C_o = \begin{cases} 1.2 \; if \; d < 0.12 \; m \; or \; if \; v_{SL} > 0.02 \; m/s \\ 2.0 \; if \; d > 0.12 \; m \; and \; if \; v_{SL} < 0.02 \; m/s \end{cases} \quad (3)$$

increases with an increase in the gas flow rate. Hasan et al.[24] reported that a transition to slug flow is expected at a void fraction of 0.25. By use of a drift-flux concept, the transition then can be expressed by Eq. 2.

The terminal rise velocity of small bubbles given by

$$v_s = 1.53\left[\frac{g\sigma_L(\rho_L - \rho_g)}{\rho_L^2}\right]^{1/4} \quad (4)$$

The Harmathy[25] expression, Eq. 4, is used for the slip or bubble-rise velocity. Transition to slug flow takes place at superficial gas velocities greater than that given by Eq. 2.

**Bubble Flow.** In bubble flows the expression for holdup, $H_L$ is

$$H_L = 1 - \frac{v_{Sg}}{C_o v_m + v_s} \quad (5)$$

where $C_0$ and $v_s$ are given by Eq. 3 and 4, respectively.

**Slug Flow.** The drift-flux model of Eq. 5 also was applied in slug flow, but with different values for $C_0$ and $v_s$ given by $C_0 = 1.2$ and

$$v_s = 0.35\sqrt{gd\frac{(\rho_L - \rho_g)}{\rho_L}}\sqrt{\sin\theta}(1+\cos\theta)^{1.2} \quad (6)$$

## 1.2 Flow model for tubing/casing annulus.

**Bubbly Flow.** For bubbly flow in an oilwell annulus, Hasan and Kabir propose that the void fraction may be estimated by an expression of the form given by:

$$f_g = \frac{v_{gs}}{Av_{gs} + B_0 v_\infty} \quad (7)$$

For gas bubbling through an annulus, the values of the parameters $A$ and $B_0$ are likely to depend on the ID and OD. A to be linearly dependent on the ID-to-OD ratio:

$$A = A_0 + A_1(d_t/d_c) \quad (8)$$

Eq. 8, of course, needs to be verified by experimental work.

With the values of $A_0$, $A_1$ and $B_0$ determined from the experimental data, the proposed model for gas void fraction during bubbly flow becomes and Using the Harmathy correlation with a constant of 1.50 to represent the effect of liquid properties on $v_\infty$ and allowing for the liquid superficial velocity, $v_{ls}$, Eq. 7 becomes

$$f_g = \frac{v_{gs}}{(1.97 + 0.371d_t/d_c)(v_{gs} + v_{ls})} + \frac{v_{gs}}{1.5[g\sigma(\rho_l - \rho_g)/\rho_l^2]^{0.25}}$$

**Slug Flow.** The model for gas void fraction in slug flow becomes

$$f_g = \frac{v_{gs}}{(1.82 + 0.9d_t/d_c)(v_{gs} + v_{ls})} + \frac{v_{gs}}{[0.30 + 0.22d_t/d_c]\sqrt{g(d_t - d_c)(\rho_l - \rho_g)/\rho_l}}$$

**Bubbly/Slug-Flow Transition.** Transition from bubbly to slug flow apparently occurs at a void fraction of about 0.25[33,34]. The validity of this transition criterion, however, remains to be established for an annular geometry. In their work, Taitel et al assumed that at the point of transition, slip ($v_g$-$v_l$) between the phases is equal to the terminal rise velocity, $v_\infty$. Because of the existence of a velocity and concentration profile, this assumption may be inappropriate. Instead, use of Eq. 7 or its equivalent can be made for a nonstagnant liquid column:

$$f_g = \frac{v_{gs}}{A(v_{gs} + v_{ls}) + B_0 v_\infty} \quad (9)$$

Because Eq. 9 is expected to apply to the entire bubbly flow regime, we can equate $f_g$=0.25 at the point of transition:

$$0.25 = \frac{v_{gs}}{A(v_{gs}+v_{ls})+B_0 v_\infty}$$

or

$$v_{gs} = \frac{1}{1-0.25A}(0.25Av_{ls}+0.25B_0 v_\infty)$$

With the appropriate expressions for the parameters $A$ and $B$, transition equation may be written as

$$v_{gs} = \frac{1}{1-0.25(1.97+0.371d_t/d_c)} \times$$
$$\times \left\{0.25(1.97+0.371d_t/d_c)v_{ls} + 0.375[g\sigma(\rho_l-\rho_g)/\rho_l^2]^{0.25}\right\}$$

### 1.3 Pressure drop calculation.

In bubble flow the total pressure gradient can be written as the sum of the gravitation or hydrostatic head $(dp/dD)_{el}$, friction $(dp/dD)_f$ and acceleration $(dp/dD)_{acc}$ components. Thus,

$$\left(\frac{dp}{dD}\right)_l = \left(\frac{dp}{dD}\right)_{el} + \left(\frac{dp}{dD}\right)_f + \left(\frac{dp}{dD}\right)_{acc} =$$
$$= \rho_s g \Delta D \sin\theta + \frac{fv_m^2 \rho_s}{2d} + v_m \rho_s \frac{dv_m}{dD} \quad (10)$$

where

$$f_C = 0.046\left(\frac{\rho_g v_{Sg} d}{\mu_g}\right)^{-0.2}(1+75\lambda_{LC})$$

$$\lambda_{LC} = \frac{q_L}{q_L - q_g}$$

$$v_m = \frac{q_L + q_g}{A} = v_{SL} + v_{Sg}$$

To estimate total pressure gradient, Eq. 10 can be used with the mixture density calculated from the liquid holdup estimated with Eq. 5. The friction component can be computed by treating the multiphase mixture as a homogeneous fluid. Friction factor, $f$, can be determined from the Moody diagram for a Reynolds number defined as

$$N_{Re\,m} = \frac{\rho_L v_m}{\mu_L}$$

This was recommended by Govier and Aziz because would not be too different from $p_m/f_{tm}$, and the contribution of the friction component to the total pressure gradient is very small.

In slug flow, as in bubble flow, the total pressure gradient can be obtained by Eq. 10 by use of Eqs. 5,6. The estimation of the friction component presents some difficulty because some of the liquid flows downward in a film around the Taylor bubble while most of the liquid flows upward in the liquid slug. Wallis[35] suggested that the wall shear stress around the vapor bubble be ignored. With this assumption, the friction pressure gradient becomes

$$\left(\frac{dp}{dD}\right)_f = \frac{f_C v_m^2 \rho_L H_L}{2d}$$

The product $p_L H_L$ is very nearly equal to pm for low-pressure systems, indicating the similarity in evaluating the friction terms in slug and bubble flow. The contribution of the friction component is no longer negligible but is still small (typically 10% of the total gradient). Acceleration, however, is small and can be neglected.

In general, the acceleration component can be neglected during all but the annular-flow pattern. Suggests that an accurate estimation of the liquid holdup is essential when computing the elevation component. This component accounts for most of the pressure drop occurring in the bubble- and slug-flow patterns. Because of the different hydrodynamics in each flow pattern, estimations of holdup, $H_L$, in-situ mixture density, $\rho_m$, and friction factor, $f$, are made separately.

## 2 BHP Calculation

Recalculation of pressure was made on each step on depth; the temperature undertook average for each of three intervals: casing, tubing/casing annulus below fluid level and tubing/casing annulus above fluid level.

### 2.1 Design procedure of calculation

Process of calculation BHP occurs in two stages:

I. The first (direct) step in the solution of this problem is the calculation of the annulus oil level (dynamic level) depth $D_l$ as a function of the bottom hole pressure $P_{wf}$ (see, for example, curve $A$ in fig.2)

II. At the second (reverse) step we use the calculated curve $D_l = f(P_{wf})$ to estimate the bottomhole pressure $P_{wfl}$ corresponding to the measured dynamic level $D_{lm}$.

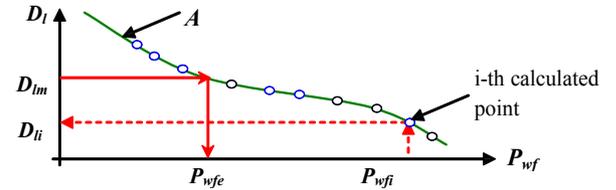

Figure 2. – The procedure of bottomhole pressure estimation.

Calculation procedure (direct step):
1. Calculate the average temperatures $\bar{T}_1, \bar{T}_2, \bar{T}_3$:

2. Calculate the bubblepoint pressure average values in the casing $(T = \bar{T}_1)$ and in the annulus $(T = \bar{T}_2)$:

$$P_b(T) = P_{bR} 10^{0.00164(T-T_R)}$$

3. Calculate the annulus gas pressure at the measured dynamic level $D_{lm}$:

$$P_l = P_a \exp\left[\frac{0.0342 D_l \gamma_g}{\bar{T}_3 \bar{z}_3}\right]$$

The value of $\bar{z}_3$ is calculated by the trial-and-error procedure:

$$\begin{array}{l} \bar{z}_3 = 1 \\ P_l = P_1 \exp\left[\dfrac{0.0342 D_l \gamma_g}{\bar{T}_3 \bar{z}_3}\right] \\ \bar{P}_3 = \dfrac{(P_1 + P_l)}{2} \\ \bar{z}_3 = f_z(\bar{P}_3, \bar{T}_3) \end{array}$$

where $f_z(P,T) =$ gas compressibility as a function of pressure $P$ and temperature $T$.

4. Determine the bottom hole pressure values to be set during calculations:

$$P_{wfi} = P_{wf\min} + \frac{(P_{wf\max} - P_{wf\min})}{N-1}(i-1)$$

where $i = 1, 2, ..., N$, $N$ – the number of calculations needed to construct the function $D_l = f(P_{wf})$ (usually N≈20).

For each value of $P_{wfi}$ $(i = 1, 2, ..., N)$ the following procedure is used to calculate the corresponding value of dynamic level depth $D_{li}$.

a) Select a depth increment $\Delta D$:
$$\Delta D = \frac{(D_w - D_p)}{N_p}$$

The value of $\Delta D$ should not be greater than 10 m.

b) Starting with known pressure value, $P_{wfi}$, at depth $D_w$, calculate a pressure traverse by the iterative procedure

$$\widetilde{P}_{j+1} = P_j - \left(\frac{dp}{dD}\right)_l^j \qquad P_{j+\frac{1}{2}} = \frac{(P_j + \widetilde{P}_{j+1})}{2}$$

$$P_{j+1} = P_j - D\left(\frac{dp}{dD}\right)_l^{j+\frac{1}{2}} \qquad D_j = D_w - j\Delta D$$

where $j = 0, 1, 2, ..., N_p$, $P_0 = P_{wfi}$, $T = \bar{T}_1$.

c) Calculate at the pump intake ($j = N_p$): $P_{up}$, $w_{gup}$, $f_{gup}$, $q_{oup}$, $\upsilon_{\sup}$.
Then estimate:
    a. separation coefficient $E_s$,
    b. annulus gas mass rate $w_{ga}$.

d) Calculate a pressure traverse in the annulus by the iterative procedure (b), where now
$D_j = D_p - j\Delta D$, $j = 0,1,2,....,$ $P_o = P_{up}$, $T = \bar{T}_2$, $\Delta D$ is to be selected properly (usually $\Delta D = 5$ m or 10 m). The iterative procedure is to be stopped when $P_{j+1} \leq P_l$.

e) Interpolate between the last two values of P to obtain the depth $D_{li}$ corresponding to pressure $P_l$:

$$D_{li} = D_j + \frac{P_j - P_l}{P_j - P_{j+1}} \Delta D$$

where $D_j$ the depth where the iteration was stopped.

Addition calculations are presented into Appendix 2.

### 2.2 Data collection

For the purpose of this study, data were collected from technology regimes on Priobskoye field. These files are formed every month and contain the information on an operating mode of all producing wells.

As candidates for BHP calculation wells which have been chosen are equipped with pressure gauges on the pump intake.

The following data were selected for calculation:
- depth (well, setting pump and fluid dynamic level);
- volumetric flow rate (oil and water);
- physical and chemical properties of fluids (oil and gas specific gravity, oil viscocity);
- casing and tubing diameter.

### Conclusions

The procedure of BHP calculation from dynamic fluid level for pumping wells completed without packer was developed during performance of the given project.

The main advantage of this method that it allows to calculate BHP without measuring dp/dt at surface as it is done in standard methods of BHP calculation. Hence allows to simplify measurements on well and to calculate BHP at any moment of production. Also this method allows to take into account real geometry of the well.

For a practical substantiation of a technique bottomhole pressure and corresponding pressure on the pump intake has been calculated on 85 wells of Priobskoye field. Results of comparison of the calculated and measured pressures are shown in figure 3. Comparison shows good accuracy of this technique ($R^2$=0.76).

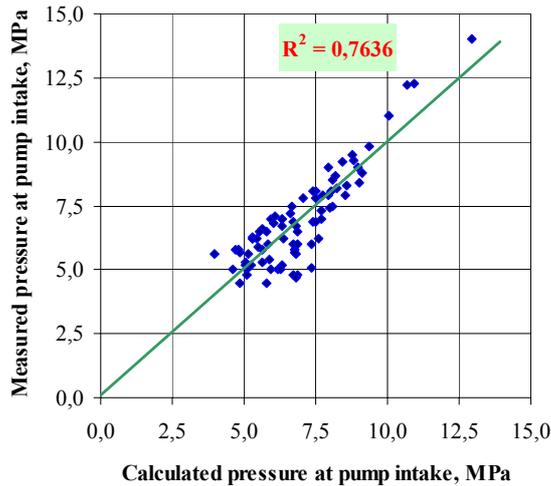

Figure 3. – Calculated and Measured pressure at intake comparison.

**Purpose of further work**

It is necessary to specify possible directions of further work for project improvement:
- use in calculations of real distribution of temperature on a well;
- creation of more exact calculation model for quantity of separated gas on the pump intake;
- specification of used correlations for PVT properties

**Nomenclature**

| | |
|---|---|
| $A$ | flow area of conduit, m$^2$, |
| $A_a$ | annular cross section area, m$^2$ |
| $A_c$ | casing inside cross section area, m$^2$ |
| $A, A_s$ | parameters, dimensionless |
| $A_{s0}, A_{s1}$ | parameters, dimensionless |
| $A_0, A_1$ | parameters, dimensionless |
| $B_o$ | oil FVF at pressure $P$ and temperature $T$, m$^3$/m$^3$ |
| $B_0$ | parameter, dimensionless |
| $B, B_s,$ | parameters, m/s |
| $C_o$ | oil isothermal compressibility, MPa$^{-1}$ |
| $C_0$ | parameter, dimensionless |
| $C_1$ | parameter, m/s |
| $d_c, d_t$ | casing ID and tubing OD, m |
| $d_{eq}$ | equivalent diameter, m |
| $D_{li}$ | dynamic level corresponding to $P_{wi}$ (calculated), m |
| $D_{lm}$ | the measured dynamic level, m |
| $D_p$ | pump setting depth, m |
| $D_w$ | bottomhole depth, m |
| $\Delta D$ | depth increment, m |
| $E_s$ | separation coefficient, dimensionless |
| $F_{gc}$ | gradient correction factor, dimensionless |
| $F_{wo}$ | water-oil ratio, dimensionless |
| $f$ | friction factor, dimensionless |
| $f_g$ | gas fraction, dimensionless |
| $f_{g\ up}$ | gas fraction at pump intake, dimensionless |
| $g$ | gravitational acceleration constant, m/s$^2$ |
| $g_T$ | temperature gradient, °K/m |
| $H_l$ | liquid holdup, dimensionless |
| $\mu_{or}$ | oil viscosity at reservoir conditions, mPa·s |
| $N_p$ | the number of depth increments, dimensionless |
| $P$ | pressure, MPa |
| $P_a$ | annulus gas pressure at the wellhead, MPa |
| $P_b$ | bubblepoint pressure at temperature $T$, MPa |
| $P_{b1}$ | bubblepoint pressure at temperature $\overline{T}_1$, MPa |
| $P_{bR}$ | bubblepoint pressure at reservoir temperature, MPa |
| $P_l$ | gas pressure at the dynamic level, MPa |
| $P_{pc}$ | pseudocritical pressure, MPa |
| $P_{sc}$ | pressure at standard conditions, MPa |
| $P_{up}$ | pump intake pressure, MPa, |
| $P_{wfe}$ | the estimated bottom hole pressure, MPa |
| $P_{wfi}$ | $i$-th value of the bottom hole pressure (set), MPa |
| $\Delta p$ | pressure increment, MPa |
| $\overline{P}_3$ | average pressure in the annulus gas cap, MPa |
| $q_t$ | total flow rate, m$^3$/day, |
| $q_{ga}$ | gas volumetric flow rate in the annulus, m$^3$/day |
| $q_o, q_w, q_g$ | flow rates (volumetric) of oil, water and gas at given $P$ and $T$, m$^3$/day |
| $q_{osc}$ | oil (volumetric) flow rate at standard conditions, m$^3$/day |
| $q_{up}$ | oil flow rate at pump intake, m$^3$/day |
| $\rho_g$ | gas density at $P$ and $T$, kg/m$^3$ |
| $\rho_L$ | liquid density at given pressure $P$ and temperature $T$ |
| $\rho_o$ | oil density at $P$ and $T$, kg/m$^3$ |
| $\rho_w$ | water density at $P$ and $T$, kg/m$^3$ |
| $R_{si}$ | initial solution GOR, m$^3$/m$^3$, dimensionless |
| $R_s(P,T)$ | solution gas/oil ratio (GOR) at given $P$ and $T$, m$^3$/m$^3$ |
| $N_{Re\ m}$ | Reynolds number of mixture, dimensionless |
| $T$ | temperature, °K |
| $\overline{T}_1$ | average temperature in casing below the pump intake, °K |
| $\overline{T}_2$ | average temperature of the oil-gas mixture in the annulus, °K |
| $\overline{T}_3$ | average temperature of the annulus gas cap, °K |
| $T_{pc}$ | pseudocritical temperature, °K |
| $T_R$ | reservoir temperature, °K |
| $T_{sc}$ | temperature at standard conditions, °K |
| $T_{wh}$ | wellhead temperature, °K |
| $v_s$ | gas slip velocity, the difference between the average gas and liquid velocities, m/s |
| $v_{s\ up}$ | gas slip velocity at pump intake, m/s |
| $v_g$ | actual gas velocity, m/s |
| $v_{sg}$ | superficial gas velocity, m/s |
| $v_l$ | actual liquid velocity, m/s |
| $v_m$ | mixture velocity, m/s |
| $v_{sl}$ | superficial liquid velocity, m/s |
| $v_{sw}$ | gas slip velocity in the water-oil mixture, m/s |
| $v_\infty$ | terminal raise velocity for a single bubble in infinite medium, m/s |

$w_g$      free gas mass flow rate in casing at pressure $P$ of interest, kg/day,
$w_{ga}$      gas mass flow rate in the annulus, kg/day,
$w_{up}$      free gas mass flow rate in the casing at suction conditions, kg/day
$\gamma_o$      oil specific gravity, dimensionless
$\gamma_g$      gas specific gravity, dimensionless
$z$      gas compressibility factor, dimensionless
$\bar{z}_3$      average gas cap compressibility, dimensionless
$\sigma$      surface tension, dynes/cm
$\theta$      inclination angle of well

**SI Metric Conversion Factors**
cP x 1.0*E-03 = Pa·s
dyne x 1.0*E-02 = mN
ft x 3.048* E-01 = m
°F(°F-32)/1.8=°C
gal x 3.785 412 E-03 = m$^3$
in. x 2.54 = cm
lbm x 4.535924E-01=kg
psi x 6.895 = kPa
°API 141.5/(131.5 + °API)     = g/cm$^3$
ft x 3.048*E-01 = m
psi x 6.895 = kPa